\documentclass[aps,prd,12pt,eqsecnum, 
preprintnumbers,nofootinbib,superscriptaddress]{revtex4}
\usepackage{amssymb,amsmath,amsthm,graphicx,amscd,color,mathtools}
\usepackage{upgreek,enumerate,color,verbatim,multirow,comment,ulem,slashed,bm}
\usepackage[hidelinks]{hyperref}

\makeatletter                    
\@addtoreset{equation}{section}  
\makeatother                     

\begin{document}

\title{Fluctuation-Dissipation Relation for Open Quantum Systems in Nonequilibrium Steady State}

\author{Jen-Tsung Hsiang}
\email{cosmology@gmail.com}
\affiliation{Center for High Energy and High Field Physics, National Central University, Chungli 32001, Taiwan}
\author{Bei-Lok Hu}
\email{blhu@umd.edu}
\affiliation{Maryland Center for Fundamental Physics and Joint Quantum Institute, University of Maryland, College Park, Maryland 20742-4111, USA}

\begin{abstract}  
Continuing our work on the nature and existence of fluctuation-dissipation relations (FDR) in linear and nonlinear open quantum systems \cite{CPR2n4D,QRad,NLFDR}, here we consider such relations when a linear system is in a nonequilibrium steady state (NESS). With the model of two-oscillators  (considered as a  short harmonic  chain with the two ends) each connected to a thermal bath of different temperatures we find that when the  chain is fully relaxed due to interaction with the baths, the relation that connects the noise kernel and the imaginary part of the dissipation kernel of the chain in one bath does not assume the conventional form for the FDR in equilibrium cases. There exists an additional term we call the `bias current' that depends on the difference of the bath's initial temperatures  and the inter-oscillator coupling strength. We further show that this term is related to the steady heat flow between the two baths when the system is in NESS. The ability to know the real time development of the inter-heat exchange (between the baths and the end-oscillators) and the intra-heat transfer (within the chain) and their dependence on the parameters in the system offers possibilities for quantifiable control and in the design of quantum heat engines or thermal devices.  
\end{abstract}


\maketitle

\baselineskip=18pt
\allowdisplaybreaks

\section{Introduction}

In an open quantum system~\cite{BrePet} description of particle-field interactions, in which the quantum system, represented by a particle or (the internal degrees of freedom) of an atom/detector, under the influence of its quantum field environment, follows a dissipative stochastic dynamics, where the noise in the environment can be  identified and the dissipative dynamics of the open system  derived. Of importance is that the backreaction of the environment on the system is included in the treatment in a self-consistent manner.  Taking this nonequilibrium statistical/stochastic mechanical  perspective~\cite{CalHu08} has many advantages. For example, it naturally shows the range of applicability of an effective field theory (EFT):  how effective a theory is is measured by the magnitude of the noise in the environment compared to the system (the usual division is taking the low energy sector as the system and the high energy sector as the environment) at the threshold when the backreaction  becomes significant -- the smaller the system is, the more effective the EFT is in offering a good description of the open system~\cite{CalHu97}.  And, because the interplay between the system and its field environment is dynamically tracked throughout its real time evolution (the imaginary-time formulation customarily used in a finite temperature field theory~\cite{FetWal,KadBay} has no place here as it is restricted to equilibrium conditions with a well defined concept of temperature), with  backreaction fully accounted, there is no free parameter in the description of the system-environment dynamics. (In fact, any coarse-graining measure introduced to the environment need be spelled out explicitly, and one can see how  different measures  give rise to different results \cite{BehHu2}). The fluctuation-dissipation relations (FDR) \cite{FordFDT} are at the heart of open quantum systems precisely because they are the exhibitions and guardians of this self-consistency.

For this reason fluctuation-dissipation relations, though rooted in statistical mechanics \cite{CalWel,Kubo,FetWal}, has wide-ranging implications and applications. For example, Sciama \cite{Sciama} treated black holes with Hawking radiation \cite{Haw} as a quantum dissipative system, and, with Candelas, proposed to view its interaction with a quantum field in the light of an FDR \cite{CadSci} (see also \cite{Mot}). Hu, Verdaguer and their co-authors \cite{HuSin95,CamVer,CamHu} showed how the backreaction of particle creation on the geometrodynamics of the early universe can be phrased in terms of an FDR.

This letter reports on the nature of an FDR for quantum systems in nonequilibrium steady state (NESS) and demonstrates its existence from the real time dynamics of a model system, that of two coupled oscillators (considered as a  short harmonic  chain with the two ends) each connected to a thermal bath of different temperatures. Quantum energy transport in this set up has been investigated by us earlier \cite{HHAoP} using an open system conceptual framework and the influence functional formalism \cite{IF,CTP}. For classical many-body systems the existence and uniqueness of the NESS is a fundamental subject and a main theme of research by mathematical physicists in statistical mechanics for decades. For Gaussian systems (such as a chain of harmonic oscillators with two heat baths at the two ends of the chain) \cite{NESShar} and anharmonic oscillators under general conditions \cite{NESSanh} there are definitive answers in the form of proven theorems. Answering this question for quantum many-body systems is not so straightforward.

\subsection{$N$ atoms/detectors in a common field environment}

Before describing our work for NESS where a system interacts with two environments it is useful to review the key issues of FDR in the configuration of one or $N$ oscillators in a common field environment. When the (near-)equilibrium condition is lifted, the fully nonequilibrium dynamics is very rich. Our approach involves calculating the evolution of the system in real time, tracking its approach to equilibrium, then from the rate of energy transfer from different sources between the system and its field environment, show the existence of a steady state  at late times and identify an FDR at work. Amongst the novel features listed below we will only delve into the first two points, so as to avoid possible confusion.

\begin{enumerate}

\item Difference between FDR based on linear response theory (LRT) and nonequilibrium (NEq) dynamics 

\item Difference between the Jarzynski-Crooks fluctuation theorems and the FDRs in our present study.

\item A static atom/detector in a quantum field bath: Role of quantum radiation and quantum dissipation in the power balance embodied in the FDR \cite{QRad}.

\item $N$ static atoms/detectors: FDR, internal energy, heat capacity and validity of the Third Law approaching absolute zero~\cite{QTD1}.

\item $N$ uniformly accelerated atoms detectors: Existence of the correlation-propagation relations (CPR) on the same footing as the FDR \cite{RHA,CPR2n4D}.

\item An anharmonic oscillator in one heat bath: Existence of a nonperturbative FDR for nonlinear open systems that can approach a stable equilibrium~\cite{NLFDR}.

\end{enumerate}

\subsubsection{Differences between LRT and NEq formulations}

Linear response theory (LRT) considers the situation when (i) the system of interest is prepared in a thermal state and \textit{remains in thermal equilibrium} with the thermal bath;  (ii) the system is subjected to a \textit{weak external disturbance} and then its responses  are registered. Because of the weak coupling with the bath and small deviations from the equilibrium state, the \textit{FDR in LRT is formulated in a perturbative sense}. 

By comparison, in the nonequilibrium (NEq) formulation, a) the system can start in any state. Once the initial state of the system and the properties of the bath are given, \textit{their interaction alone determines the entire evolution history} described by the reduced dynamics of the system coarse-grained over the environmental influences.~b) In the NEq context, the system and its environment can be \textit{strongly coupled while dynamically evolving}, but the existence of an equilibrium state is not a priori known or given. One needs to first determine if the system comes to equilibration with its environment, a precondition for an FDR to exist for that equilibrium state.

Therefore FDR in a NEq context is an emergent phenomenon depending on many factors which enter into the nonequilibrium dynamics of the open system. As such it is more complex as it involves dynamical relaxation of the system into equilibrium, and the existence of an FDR has a dynamical significance since it ensures the
balance of the energy flow between the reduced system and
the environment.  

\subsubsection{Difference from NEq fluctuation theorem contexts} 

A major advancement in NEq sciences occurred in the 90s stemming from the formulation of the fluctuation theorems, first in the Evans-Searle (ES)~\cite{ESfluc} and Gallavotti-Cohen (GC)~\cite{GCfluc} veins, and then in the Jarzynski-Crooks (JC)~\cite{Jarzyn,Crooks} work relations. The set up in the former, of the EV-GC veins,  is also for NESS, but the emphasis is in large deviations in the stochastic dynamics and mostly for classical systems. Whereas in the JC theorems NESS is not required and work input from external agents enters in the relations. In our set up we only consider energy flow or heat transfer, leaving out work completely -- in fact, quantum work may not be a well defined concept (see, e.g.,~\cite{SubHu}), pending further investigations. The  physics in all three situations is very rich which injects new vitality in the development of NEq sciences. 

Though not directly related to our present problem we note a few representative papers on fluctuation theorems in NESS \cite{Che,SpeSei,ProPar,Sei,Gua} for interested readers.

\subsection{Main Results of this work: FDR in NESS}
In an earlier paper \cite{HHAoP} we have explored quantum transport in the same setup  but short of showing the FDR explicitly. We do it here, with new understandings in the interplay of the inter- and intra- components of energy transport in open quantum systems in NESS. From the nonequilibrium dynamics of all the constituents in the system together with those interacting with the two baths,  we find that when  the chain is fully relaxed, the relation that connects the noise kernel and the imaginary part of the dissipation kernel of the chain in one bath does not assume the conventional form of the FDR for the system in a single bath after equilibration. There exists an additional term, a `bias current', that depends on the difference of the bath's initial temperatures  and the inter-oscillator coupling strength. We further show that this bias current is related to the steady thermal flow between the baths in the nonequilibrium steady state. Thus the ability to know the real time development of the inter-heat exchange (between the baths and the end-oscillators) and the intra-heat transfer (within the chain) and their dependence on the parameters in the system offers possibilities for quantifiable control and in the design of quantum heat engines or thermal devices. 

This paper is organized as follows:  Sec.~\ref{S:rbtru} uses a simple example of one oscillator coupled with a thermal field bath to illustrate how an FDR arises through relaxation from a nonequilibrium evolution.  Sec.~\ref{S:vkjdfd} considers two coupled oscillators representing a harmonic chain, each interacting with its own bath at two different initial temperatures. While results from the previous section describes the activities of the two end oscillators separately, the situation changes when the two baths of different temperatures are connected through a chain. When the dynamics of the harmonic chain is fully relaxed, we find a  relation that connects the noise kernel and the imaginary part of the dissipation kernel. The difference from the FDR for one oscillator with one bath is the focus of our attention in deriving an FDR for a system in NESS and understanding its physical meaning. In Sec.~\ref{S:pturgf}, we conclude with some general remarks on FDR in NESS.
  

\section{Equilibration and FDR at the thermal bath junctions}\label{S:rbtru}

We divide our analysis into two parts: i) what happens at the junctions in terms of heat transfer from the bath of higher temperature $T_H$ through the first oscillator of the chain. What happens at the other end of the chain is just the reverse. See \cite{HHAoP} for details. ii) what happens in the chain between the two end oscillators.  We shall discuss i) in the section and ii) in the next section. 

To see what happens at the bath-oscillator junctures and to gain some insight into the contents and the meanings of the fluctuation-dissipation relation (FDR) it is advantageous to take a quantum open systems viewpoint and derive the nonequilibrium dynamics of the system.  We give a short description of this endeavor, with physical explanations of what enters into this relation. A more general treatment is given in~\cite{QTD1}. 

We consider a system of just one  harmonic oscillator interacting with a thermal quantum field bath. Let $\hat{\chi}$ denote the displacement operator of this oscillator which could be used to represent the internal degrees of freedom (idf) of an atom (a harmonic atom) or a detector (an Unruh-DeWitt `detector' -- a common terminology used in the quantum field in curved spacetime and relativistic quantum information communities. We shall use these two terms  `atom' and `detector' interchangeably while reserving `oscillator' for the idf.) The equations of motion of $\hat{\chi}(t)$ and the bath field operator $\hat{\phi}(t,\mathbf{x})$ are given respectively by
\begin{align}
	\ddot{\hat{\chi}}(t)+\omega^{2}_{\textsc{b}}\,\hat{\chi}(t)&=\frac{e}{m}\,\hat{\phi}(t,\mathbf{z})\,,\label{E:fghjfdf}\\
	\partial^{2}_{t}\hat{\phi}(t,\mathbf{x})-\bm{\nabla}^{2}\hat{\phi}(t,\mathbf{x})&=e\,\hat{\chi}(t)\,\delta^{(3)}(\mathbf{x}-\mathbf{z})\,,\label{E:ghnfbs}
\end{align}
where the oscillator's mass is $m$ and bare natural frequency $\omega_{\textsc{b}}$, while the coupling strength $e$ between the oscillator and the bath field is not restricted to a small value. The Cartesian coordinate of the Minkowski spacetime is generically denoted by $x^{\mu}=(t,\mathbf{x})$. The formal solution to \eqref{E:ghnfbs}
\begin{equation}\label{E:ghrhfgsf}
	\hat{\phi}(t,\mathbf{x})=\hat{\phi}_{h}(t,\mathbf{x})+e\int\!d^{4}x'\;G_{R,0}^{(\phi)}(x,x')\,\hat{\chi}(t')\,\delta^{(3)}(\mathbf{x}-\mathbf{z})
\end{equation}
upon substituting into \eqref{E:fghjfdf} gives a reduced description of the oscillator, in the form of a quantum Langevin equation
\begin{equation}\label{E:hhdsmms}
	\ddot{\hat{\chi}}(t)+\omega^{2}_{\textsc{b}}\,\hat{\chi}(t)=\frac{e}{m}\,\hat{\phi}_{h}(t,\mathbf{z})+\frac{e^{2}}{m}\int\!d^{4}x'\;G_{R,0}^{(\phi)}(t,\mathbf{z};t',\mathbf{z})\,\hat{\chi}(t')\,,
\end{equation}
where $x=(t,\mathbf{x})$, $x'=(t',\mathbf{x}')$ and $G_{R,0}^{(\phi)}(x,x')$ is the retarded Green's function of the free field, defined by
\begin{equation}
	G_{R,0}^{(\phi)}(x,x')=i\,\theta(t-t')\,\bigl[\hat{\phi}_{h}(x),\hat{\phi}_{h}(x')\bigr]\,,
\end{equation}
which by construction is independent of the field state. .  Eq.~\eqref{E:ghrhfgsf} then tells that a radiation field is emitted from the atom  at the position $\mathbf{z}$ at an earlier time $t'$ as a consequence of the interaction with the atom, and is superposed onto the original free field, described by the homogeneous solution $\hat{\phi}_{h}(t,\mathbf{x})$ of \eqref{E:ghnfbs}.

The first term on the right hand side of \eqref{E:hhdsmms} is the stochastic forcing term associated with the  quantum fluctuations of the free field. This ubiquitous `noise force'  imparts a stochastic component into the oscillator's motion. The second  term involves the coincident limit of the Green's function of the massless field and thus needs regularization; the cutoff-dependent part will regularize the bare frequency to its physical value and the remaining finite part  describes the reaction to the radiation field which gives rise to a frictional `self force'. Thus the reduced equation of motion \eqref{E:hhdsmms} becomes
\begin{equation}\label{E:fkddre}
	\ddot{\hat{\chi}}(t)+2\gamma\,\dot{\hat{\chi}}(t)+\omega^{2}_{\textsc{r}}\,\hat{\chi}(t)=\frac{e}{m}\,\hat{\phi}_{h}(t,\mathbf{z})
\end{equation}
where $\omega_{\textsc{r}}$ represents the physical  frequency and $\gamma=e^{2}/8\pi m$ is the damping constant. The second term on the left hand side describes the self-force. The noise force and the self-force  compete with and balance off each other: on the one hand, the noise force imparts energy of the field into the oscillator while the self-force drains the oscillator's energy back to the field environment. They will account for the energy exchange between the oscillator and its environment. In electromagnetism  it is the atom's idf responding to the vacuum fluctuations of the field in the form of emitted radiation and its reaction force;  in thermodynamics, it is the harmonic oscillator in the form of heat transfer and back-action. It can be shown that when the dynamics of the idf of the detector is fully relaxed there is a balance between these two forces, $\xi$ denoting the noise from the  fluctuations of the quantum field and $\gamma$ denoting the damping of the reactive self-force. That is, there is no net energy flow in either direction, or, the total power in the system-environment exchange vanishes.   
\begin{equation}\label{E:fxvjhe}
	\lim_{t\gg\gamma^{-1}}P_{\textsc{h}}(t)=\lim_{t\gg\gamma^{-1}}\Bigl[P_{\xi}(t)+P_{\gamma}(t)\Bigr]=0\,,
\end{equation}
if we define the powers delivered by the noise force and the damping force respectively by
\begin{align}\label{E:dkhbger}
	P_{\xi}(t)&=\frac{1}{2}\langle\bigl\{e\,\hat{\phi}_{h}(t,\mathbf{x}),\dot{\hat{\chi}}(t)\bigr\}\rangle\,,&P_{\gamma}(t)&=-\frac{1}{2}\langle\bigl\{2m\gamma\,\dot{\hat{\chi}}(t),\dot{\hat{\chi}}(t)\bigr\}\rangle\,,
\end{align}
which is the power expression in Newtonian mechanics arranged in a symmetric operator ordering. The expectation values in \eqref{E:dkhbger} is taken with respect to the initial state of the scalar field, since we are concerned only with late-time dynamics. The contribution related to the initial configuration of the oscillator will become exponentially small at times greater than the relaxation time scale $\gamma^{-1}$, so it is discarded in this context.

What does this equilibration condition at late times say  about the relations between the Green's functions in the field and the system? In the form of the retarded Green's functions $G_{R}^{(\chi)}$, the Hadamard functions $G_{H}^{(\chi)}$ of the interacting oscillator, described by \eqref{E:fkddre}, and the counterparts $G^{(\phi)}_{R,\,0}$,  $G^{(\phi)}_{H,\,\beta}$ of the free quantum field in its initial thermal state, the net energy flow $P_{\textsc{h}}(t)$ can be expressed as
\begin{equation}
	P_{\textsc{h}}(t)=e^{2}\int_{0}^{t}\!ds\;\biggl\{\frac{d}{dt}G_{R}^{(\chi)}(t,s)\,G_{H,\,\beta}^{(\phi)}(s,t)-\Gamma_{R,\,0}^{(\phi)}(t-s)\frac{d^{2}}{ds\,dt}G_{H}^{(\chi)}(s,t)\biggr\}\,,
\end{equation}
where 
\begin{equation}\label{E:ithrhkfd}
	G_{R,\,0}^{(\phi)}(t,\mathbf{z};s,\mathbf{z})\equiv G_{R,\,0}^{(\phi)}(t-s)=\frac{d}{ds}\Gamma_{R,\,0}^{(\phi)}(t-s)\,.
\end{equation}
Then in the limit $t\to\infty$, it can be shown that 
\begin{equation}\label{E:rsdjwe}
	\lim_{t\to\infty}P_{\textsc{h}}(t)=\int_{-\infty}^{\infty}\!\frac{d\kappa}{2\pi}\;\kappa\,\biggl\{\coth\frac{\beta\kappa}{2}\,\operatorname{Im}\widetilde{G}_{R}^{(\chi)}\!(\kappa)-\widetilde{G}_{H}^{(\chi)}(\kappa)\biggr\}\,\operatorname{Im}\widetilde{G}_{R,\,0}^{(\phi)}(\kappa)\,.
\end{equation}
We have used the FDR of the free scalar field associated with its initial thermal state at temperature $\beta^{-1}$
\begin{equation}\label{E:rhfgjdh}
	\widetilde{G}_{H,\,\beta}^{(\phi)}(\kappa)=\coth\frac{\beta\kappa}{2}\,\operatorname{Im}\widetilde{G}_{R,\,0}^{(\phi)}(\kappa)\,.
\end{equation}
in deriving \eqref{E:rsdjwe}, and defined the Fourier transformation of a function $f(t)$ by
\begin{equation}
	\widetilde{f}(\kappa)=\int_{-\infty}^{\infty}\!dt\;f(t)\,e^{+i\kappa t}\,.
\end{equation}
The balance of the energy flow \eqref{E:fxvjhe} at late times then implies an FDR for the interacting harmonic oscillator
\begin{equation}\label{E:djgeresfjk}
	\widetilde{G}_{H}^{(\chi)}(\kappa)=\coth\frac{\beta\kappa}{2}\,\operatorname{Im}\widetilde{G}_{R}^{(\chi)}\!(\kappa)\,.
\end{equation}
The derivation clearly indicates the connection between equilibration, energy balance, and the FDR. Note that in the above discussion, the state of the oscillator in general can be quite different from the initial state, as well as the final equilibrium state. Here lies one of many important differences between our NEq dynamics approach and the linear response theory. In addition, Eqs.~\eqref{E:fkddre} and~\eqref{E:fxvjhe} imply
\begin{equation}\label{E:fguiwow}
	\frac{d}{dt}\Bigl[\frac{m}{2}\,\dot{\chi}^{2}(t)+\frac{m\omega_{\textsc{r}}}{2}\,\hat{\chi}^{2}(t)\Bigr]=P_{\textsc{h}}(t)\to0
\end{equation}
for $t\gg\gamma^{-1}$. This result is stronger than what \eqref{E:fkddre} tells at late times. The latter, that is, energy conservation, only gives
\begin{equation}
	\frac{d}{dt}\Bigl[\frac{m}{2}\,\dot{\chi}^{2}(t)+\frac{m\omega_{\textsc{r}}}{2}\,\hat{\chi}^{2}(t)\Bigr]=P_{\textsc{h}}(t)\,.
\end{equation}
Furthermore, from \eqref{E:fguiwow} we see that after relaxation, the oscillator will act like a free harmonic oscillator with the renormalized frequency $\omega_{\textsc{r}}$, following a reversible dynamics, and obeying the FDR \eqref{E:djgeresfjk}. However, note that the reduced density matrix of this relaxed oscillator does not take on the Gibbs form. That is, it is not a thermal state unless the oscillator-field coupling is vanishingly small.  This latter condition is a tacit yet pivotal assumption in the underpinnings of equilibrium statistical thermodynamics, manifested here as a precondition for the establishment of the canonical ensemble. 


\section{Bias Current and FDR in System under NESS}\label{S:vkjdfd}

Now we take what we have learned between one oscillator and its bath as happening at both ends of a harmonic chain, our system, and focus on the dynamics of the coupled oscillators interacting with two baths at the two ends. If this system can reach a steady state we shall be able to determine whether an FDR exists and the role it plays in NESS. We consider the case of two coupled oscillators for simplicity, without sacrifice of the physics we seek after -- extension to an $N$-oscillator chain is straightforward~\cite{HHAoP}. Each oscillator has its own private bath,  modeled by a massless scalar field bilinearly coupled to it. Initially both baths are uncorrelated and prepared in their individual thermal states at different initial temperatures. The coupling between these two oscillators will bring together the influence of each oscillator's private bath. It is the linkage between what we learned earlier and what we are to explore presently. 

As a transition to the general FDR in NESS discussion, let us take a look at two special cases: 1) Zero inter-oscillator coupling. This severs the two oscillators and what each oscillator does with its own bath is ab initio independent of the other. Conclusions from the previous section will apply to both: the two end oscillators will enjoy an FDR of different initial bath temperatures. 2) When both private baths have the same temperature. With zero temperature gradient there will be no thermal energy flux through the chain between two baths. Thus, at least from the viewpoint of the averaged energy flow, each oscillator acts independently and does not affect one another. Each has its own FDR with the same temperature parameter. This still holds even though each oscillator may have a different coupling strength with its private bath, because  for  bilinear oscillator-bath coupling the coupling strength does not enter in the FDR. (Note the vanishing of such a thermal flow on average does not necessarily imply there is no fluctuations of energy flow in this case.)  What makes this possible is because the whole system is in equilibrium, and as we saw earlier, under this condition the oscillators are effectively set `free' and thus behave as if they are independent of each other. 

Let us now consider the situation when the two initial bath temperatures are different, and the  inter-oscillator coupling is nonvanishing. The moment we make the two temperatures different heat will begin to flow from the high temperature bath through the chain unto the low temperature bath. In so doing the equilibrium condition is nullified, and the FDR (at least in the form discussed in the previous section) for each of the end oscillators with its own bath, which is predicated upon the existence of an equilibrium condition, no longer exists. This writes off the activities at the two ends and our attention will be shifted to the heat flow in the chain between the two end oscillators.  Thus we can reason that if there exists an FDR for the system it must be connected to the behavior of this heat flow, which we call the `bias current'. We shall show  that indeed this is the case: if we can prove that a NESS exists at late times for the system, the FDR is embedded in this bias current. 

The equations of motion for the two oscillators in this context are
\begin{align}
	\ddot{\hat{\chi}}_{1}(t)+\omega_{1\textsc{b}}^{2}\,\hat{\chi}_{1}(t)-\frac{e_{1}^{2}}{m}\int_{0}^{t}\!ds\;G_{R,0}^{(\phi_{1})}(t-s)\,\hat{\chi}_{1}(s)+\sigma\,\hat{\chi}_{2}(t)&=\frac{e_{1}}{m}\,\hat{\phi}_{1h}(t)\,,\\
	\ddot{\hat{\chi}}_{2}(t)+\omega_{2\textsc{b}}^{2}\,\hat{\chi}_{2}(t)-\frac{e_{2}^{2}}{m}\int_{0}^{t}\!ds\;G_{R,0}^{(\phi_{2})}(t-s)\,\hat{\chi}_{2}(s)+\sigma\,\hat{\chi}_{1}(t)&=\frac{e_{2}}{m}\,\hat{\phi}_{2h}(t)\,.
\end{align}
where $\sigma$ is the strength of the inter-oscillator coupling.  The operator $\hat{\chi}_{i}$ represents the displacement of the $i^{\text{th}}$ oscillator, whose bare oscillating frequency is $\omega_{i\textsc{b}}$, and $e_{i}$ is the coupling strength with its private bath field $\hat{\phi}_{i}$. We assume that both oscillators have the same mass $m$. The interpretation of these equations of motion is similar to that associated with \eqref{E:hhdsmms}.

It is convenient to write them into a compact matrix form
\begin{equation}\label{E:gbtyer}
	\ddot{\mathbf{X}}(t)+\bm{\Omega}_{\textsc{b}}^{2}\cdot\mathbf{X}(t)-\frac{1}{m}\int_{0}^{t}\!ds\;\mathbf{C}\cdot\mathbf{G}_{R,0}^{(\phi)}(t-s)\cdot\mathbf{C}\cdot\mathbf{X}(s)=\frac{1}{m}\,\mathbf{C}\cdot\bm{\Phi}_{h}(t)\,,
\end{equation}
with
\begin{align*}
	\mathbf{X}(t)&=\begin{pmatrix}\hat{\phi}_{1}(t)\\\hat{\phi}_{2}(t)\end{pmatrix}\,,\qquad\qquad\mathbf{C}=\begin{pmatrix}e_{1} &0\\0 &e_{2}\end{pmatrix}\,, \qquad\qquad\bm{\Omega}_{\textsc{b}}^{2}=\begin{pmatrix}\omega_{1\textsc{b}}^{2} &\sigma\\\sigma &\omega_{2\textsc{b}}^{2}\end{pmatrix}\,,\\
	\mathbf{G}_{R}^{(\phi)}(t-s)&=\begin{pmatrix}G_{R,0}^{(\phi_{1})}(t-s) &0\\0 &G_{R,0}^{(\phi_{2})}(t-s)\end{pmatrix}\,.
\end{align*}
Since it has be shown~\cite{HHAoP} that the steady state exists for such a configuration at late times we will focus on the late-time dynamics of \eqref{E:gbtyer}. The Fourier transformation of \eqref{E:gbtyer} gives
\begin{equation}\label{E:ytvds}
	\Bigl[-\kappa^{2}\,\mathbf{I}+\bm{\Omega}_{\textsc{b}}^{2}-\frac{1}{m}\,\mathbf{C}\cdot\widetilde{\mathbf{G}}_{R,0}^{(\phi)}(\kappa)\cdot\mathbf{C}\Bigr]\cdot\widetilde{\mathbf{X}}(\kappa)=\frac{1}{m}\,\mathbf{C}\cdot\widetilde{\bm{\Phi}}_{h}(\kappa)\,,
\end{equation}
and Eq.~\eqref{E:ytvds} then gives $\widetilde{\mathbf{X}}(\kappa)=\widetilde{\mathbf{G}}_{R}^{(\chi)}(\kappa)\cdot\mathbf{C}\cdot\widetilde{\bm{\Phi}}_{h}(\kappa)$, in which the retarded Green's function matrix $\widetilde{\mathbf{G}}_{R}^{(\chi)}(\kappa)$ of the interacting oscillator in the frequency space is given by
\begin{equation}\label{E:dkbshergdf}
	\widetilde{\mathbf{G}}_{R}^{(\chi)}(\kappa)=\frac{1}{m}\,\Bigl[-\kappa^{2}\,\mathbf{I}+\bm{\Omega}_{\textsc{b}}^{2}-\frac{1}{m}\,\mathbf{C}\cdot\widetilde{\mathbf{G}}_{R,0}^{(\phi)}(\kappa)\cdot\mathbf{C}\Bigr]^{-1}\,.
\end{equation}
The complete solution to the reduced equation of motion \eqref{E:gbtyer} is then
\begin{equation}\label{E:kfkjfskd}
	\mathbf{X}(t)=\mathbf{X}_{h}(t)+\int_{0}^{t}\!\frac{d\kappa}{2\pi}\;\widetilde{\mathbf{G}}_{R}^{(\chi)}(\kappa)\cdot\mathbf{C}\cdot\widetilde{\bm{\Phi}}_{h}(\kappa)\,e^{-i\kappa t}\,,
\end{equation}
where $\mathbf{X}_{h}(t)$ is the corresponding homogeneous solution to \eqref{E:gbtyer}, depending on the initial conditions, but its form is irrelevant in the following discussion because it will decay with time. Eq.~\eqref{E:kfkjfskd} will allow us to construct the various two-point Green's functions of the interacting oscillators. For example, the Schwinger two-point functions associated with the oscillators can be found by
\begin{equation}\label{E:gbjfhdfs}
	\mathbf{G}_{>}^{(\chi)}(t,t')=i\,\langle\mathbf{X}(t)\mathbf{X}^{T}(t')\rangle\,.
\end{equation}
In general, following similar arguments presented in the previous section, the oscillator chain connecting two thermal baths will undergo nonequilibrium evolution with time. Hence the associated two-point functions will not be stationary in time. In other words,  they will not be functions of the difference of two time arguments,
\begin{equation}
	\mathbf{G}_{>}^{(\chi)}(t,t')\neq\mathbf{G}_{>}^{(\chi)}(t-t')\,.
\end{equation}
Nonetheless, it has been shown in~\cite{HHAoP,QTD1} that when both $t$ and $t'$ are far larger than the inverse of the damping constants, the nonstationary components of the two-point functions of the oscillators are exponentially suppressed. The Schwinger function  in \eqref{E:gbjfhdfs} then reduces to the form
\begin{equation}
	\mathbf{G}_{>}^{(\chi)}(t,t')=\int_{-\infty}^{\infty}\!\frac{d\kappa}{2\pi}\;\widetilde{\mathbf{G}}_{R}^{(\chi)}(\kappa)\cdot\mathbf{C}\cdot\widetilde{\mathbf{G}}_{>,0}^{(\phi)}(\kappa)\cdot\mathbf{C}\cdot\widetilde{\mathbf{G}}_{R}^{(\chi)T}(-\kappa)\,e^{-i\kappa(t-t')}\,,
\end{equation}
at late times, so that its Fourier transform is given by
\begin{align}
	\widetilde{\mathbf{G}}_{>}^{(\chi)}(\kappa)&=\widetilde{\mathbf{G}}_{R}^{(\chi)}(\kappa)\cdot\mathbf{C}\cdot\widetilde{\mathbf{G}}_{>,0}^{(\phi)}(\kappa)\cdot\mathbf{C}\cdot\widetilde{\mathbf{G}}_{R}^{(\chi)T}(-\kappa)\,,\label{E:fgbkfffhbd}
\end{align}
where $\widetilde{\mathbf{G}}_{>,0}^{(\phi)}(\kappa)$ is the Fourier transform of the Schwinger function of the free quantum field $\hat{\bm{\Phi}}_{h}$. Note that both $\widetilde{\mathbf{G}}_{R}^{(\chi)}(\kappa)$ and $\widetilde{\mathbf{G}}_{>}^{(\chi)}(\kappa)$ of the interacting oscillators are symmetric matrices.

Now we attempt to construct an FDR for the NESS configuration. Observing from \eqref{E:dkbshergdf}, we find
\begin{align}
	\widetilde{\mathbf{G}}_{R}^{(\chi)}(\kappa)-\widetilde{\mathbf{G}}_{R}^{(\chi)}(-\kappa)&=\widetilde{\mathbf{G}}_{R,0}^{(\phi)}(\kappa)\cdot\mathbf{C}\cdot\Bigl[\widetilde{\mathbf{G}}_{R,0}^{(\phi)}(\kappa)-\widetilde{\mathbf{G}}_{R,0}^{(\phi)}(-\kappa)\Bigr]\cdot\mathbf{C}\cdot\widetilde{\mathbf{G}}_{R,0}^{(\phi)}(-\kappa)\,,
\end{align}
by the operator identity
\begin{equation}
	\mathbf{A}^{-1}-\mathbf{B}^{-1}=-\mathbf{A}^{-1}\bigl(\mathbf{A}-\mathbf{B}\bigr)\mathbf{B}^{-1}\,,
\end{equation}
for any two invertible operators $\mathbf{A}$, $\mathbf{B}$. Thus we obtain
\begin{equation}\label{E:ytubds}
	\operatorname{Im}\widetilde{\mathbf{G}}_{R}^{(\chi)}(\kappa)=\widetilde{\mathbf{G}}_{R}^{(\chi)}(\kappa)\cdot\mathbf{C}\cdot\operatorname{Im}\widetilde{\mathbf{G}}_{R,0}^{(\phi)}(\kappa)\cdot\mathbf{C}\cdot\widetilde{\mathbf{G}}_{R}^{(\chi)}(-\kappa)\,.
\end{equation}
The noise kernel can be identified by
\begin{align}
	\widetilde{\mathbf{G}}_{H}^{(\chi)}(\kappa)&=-\frac{i}{2}\Bigl[\widetilde{\mathbf{G}}_{>}^{(\chi)}(\kappa)+\widetilde{\mathbf{G}}_{>}^{(\chi)T}(-\kappa)\Bigr]=\widetilde{\mathbf{G}}_{R}^{(\chi)}(\kappa)\cdot\mathbf{C}\cdot\widetilde{\mathbf{G}}_{H,0}^{(\phi)}(\kappa)\cdot\mathbf{C}\cdot\widetilde{\mathbf{G}}_{R}^{(\chi)\dagger}(\kappa)\,,\label{E:gnkdfjer}
\end{align}
at late times, because the Hadamard function $\mathbf{G}_{H}^{(\chi)}(t,t')$ can be related to the Schwinger function $\mathbf{G}_{>}^{(\chi)}(t,t')$ by
\begin{align}
	\mathbf{G}_{H}^{(\chi)}(t,t')&=-\frac{i}{2}\Bigl[\mathbf{G}_{>}^{(\chi)}(t,t')+\mathbf{G}_{>}^{(\chi)T}(t',t)\Bigr]\notag\\
	&=-\frac{i}{2}\int_{-\infty}^{\infty}\!\frac{d\kappa}{2\pi}\;\Bigl[\widetilde{\mathbf{G}}_{R}^{(\chi)}(\kappa)\cdot\mathbf{C}\cdot\widetilde{\mathbf{G}}_{>,0}^{(\phi)}(\kappa)\cdot\mathbf{C}\cdot\widetilde{\mathbf{G}}_{R}^{(\chi)\dagger}(\kappa)\Bigr.\label{E:dbkerjd}\\
	&\qquad\qquad\qquad\qquad\qquad\qquad+\Bigl.\widetilde{\mathbf{G}}_{R}^{(\chi)}(\kappa)\cdot\mathbf{C}\cdot\widetilde{\mathbf{G}}_{>,0}^{(\phi)\dagger}(\kappa)\cdot\mathbf{C}\cdot\widetilde{\mathbf{G}}_{R}^{(\chi)\dagger}(\kappa)\Bigr]\, e^{-i\kappa(t-t')}\,.\notag
\end{align}
The second equality holds only at late times. Note that in the current setup, the Green's function matrix of the free field is diagonal, such that $\mathbf{G}^{(\phi)}(t,t')=\mathbf{G}^{(\phi)T}(t,t')$ and $\widetilde{\mathbf{G}}^{(\phi)}(\kappa)=\widetilde{\mathbf{G}}^{(\phi)T}(\kappa)$.

Now we compare \eqref{E:gnkdfjer} with \eqref{E:ytubds}, and we seem to have an FDR for the oscillator. However, we notice that for the free bath fields initially in their individual thermal states of different temperatures, the noise kernel $\widetilde{\mathbf{G}}_{H,0}^{(\phi)}$ and the imaginary part of dissipation kernel $\widetilde{\mathbf{G}}_{R,0}^{(\phi)}$ obey a matrix FDR rather than a simple relation like \eqref{E:rhfgjdh},
\begin{align}
	\widetilde{\mathbf{G}}_{H,0}^{(\phi)}(\kappa)&=\widetilde{\mathbf{F}}(\kappa)\cdot\operatorname{Im}\widetilde{\mathbf{G}}_{R,0}^{(\phi)}(\kappa)=\widetilde{\mathbf{f}}(\kappa)\cdot\operatorname{Im}\widetilde{\mathbf{G}}_{R,0}^{(\phi)}(\kappa)\cdot\widetilde{\mathbf{f}}(\kappa)\,,\label{E:fkghrtkgdf}\\
\intertext{with}
	\widetilde{\mathbf{F}}(\kappa)&=\widetilde{\mathbf{f}}^{2}(\kappa)=\begin{pmatrix}\coth\dfrac{\beta_{1}\kappa}{2} &0\\[6pt]0&\coth\dfrac{\beta_{2}\kappa}{2}\end{pmatrix}\,,
\end{align}
and $\beta_{i}^{-1}$ be the initial temperature of the private bath of oscillator $i$.  That is, the kernel functions in the FDR \eqref{E:fkghrtkgdf} of the free field in the NESS configuration are not related by a scalar factor. Instead, they are connected by a diagonal matrix $\widetilde{\mathbf{F}}(\kappa)$. This will be the obstacle of writing an FDR for the oscillators into the conventional form in the NESS setting.

Let us write \eqref{E:gnkdfjer} in a form as close as possible to the conventional FDR like \eqref{E:fkghrtkgdf}. From \eqref{E:dbkerjd}, we have
\begin{align}
	\widetilde{\mathbf{G}}_{H}^{(\chi)}(\kappa)&=\widetilde{\mathbf{G}}_{R}^{(\chi)}(\kappa)\cdot\mathbf{C}\cdot\widetilde{\mathbf{F}}(\kappa)\cdot\operatorname{Im}\widetilde{\mathbf{G}}_{R,0}^{(\phi)}(\kappa)\cdot\mathbf{C}\cdot\widetilde{\mathbf{G}}_{R}^{(\chi)\dagger}(\kappa)\notag\\
	&=\widetilde{\mathbf{F}}(\kappa)\cdot\operatorname{Im}\widetilde{\mathbf{G}}_{R}^{(\chi)}(\kappa)+\bigl[\widetilde{\mathbf{G}}_{R}^{(\chi)}(\kappa),\,\widetilde{\mathbf{F}}(\kappa)\bigr]\cdot\widetilde{\mathbf{G}}_{R}^{(\chi)-1}(\kappa)\cdot\operatorname{Im}\widetilde{\mathbf{G}}_{R}^{(\chi)}(\kappa)\,,\label{E:rnkdfert}
\end{align}
where we note that the matrix $\mathbf{C}$ is diagonal and
\begin{equation}\label{E:kfgjrtw}
	\bigl[\widetilde{\mathbf{G}}_{R}^{(\chi)}(\kappa),\,\widetilde{\mathbf{F}}(\kappa)\bigr]=-\bigl(\coth\dfrac{\beta_{1}\kappa}{2}-\coth\dfrac{\beta_{2}\kappa}{2}\bigr)\,\bigl[\widetilde{\mathbf{G}}_{R}^{(\chi)}(\kappa)\bigr]_{12}\,\mathbf{J}\,,
\end{equation}
with
\begin{align}
	\mathbf{J}&=\begin{pmatrix}0 &+1\\-1 &0\end{pmatrix}\,,&\mathbf{J}^{\dagger}=\mathbf{J}^{-1}=-\mathbf{J}\,,
\end{align}
and
\begin{align}
	\widetilde{\mathbf{G}}_{R}^{(\chi)-1}(\kappa)&=\frac{1}{\det\widetilde{\mathbf{G}}_{R}^{(\chi)}(\kappa)}\,\mathbf{J}^{-1}\cdot\widetilde{\mathbf{G}}_{R}^{(\chi)}(\kappa)\cdot\mathbf{J}\,,\\
	\det\widetilde{\mathbf{G}}_{R}^{(\chi)}(\kappa)&=\widetilde{\mathbf{G}}_{R}^{(\chi)}(\kappa)\cdot\mathbf{J}^{-1}\cdot\widetilde{\mathbf{G}}_{R}^{(\chi)}(\kappa)\cdot\mathbf{J}\,
\end{align}
Thus in \eqref{E:rnkdfert}, the relation connecting the noise kernel and the dissipation kernel does not satisfy the traditional form of the FDR or as \eqref{E:fkghrtkgdf}. There is an additional term related to the temperature difference between two thermal baths, which seems to account for the heat flow between the baths. In addition, we observe that this term is proportional to the 1-2 component of the retarded Green's function matrix $\widetilde{\mathbf{G}}_{R}^{(\chi)}(\kappa)$ of the oscillators, that is, linking oscillator 1 and 2. The presence of the matrix $\mathbf{J}$, from  hindsight, reflects the asymmetry between oscillator 1 and oscillator 2. That is, if the initial temperature difference between the two private baths is fixed, then the heat current flows from bath 1 to oscillator 1 will be in the opposite direction to the heat current from bath 2 to oscillator 2.  Finally, we observe that the commutator $\bigl[\widetilde{\mathbf{G}}_{R}^{(\chi)}(\kappa),\,\widetilde{\mathbf{F}}(\kappa)\bigr]$ in \eqref{E:kfgjrtw} will vanish when either both private bath have the same initial temperature or the retarded Green's function matrix $\widetilde{\mathbf{G}}_{R}^{(\chi)}(\kappa)$ is diagonal. Both correspond to the trivial cases that there is no thermal energy flow between the two baths. Thus, \eqref{E:rnkdfert} reduces to the conventional FDRs in a matrix form, as discussed in the beginning of this section.

Next we wish to discuss the physical meaning of the additional term in \eqref{E:rnkdfert} and to find its connection with the heat current through the oscillator chain in the NESS. We will examine the energy flows between  bath 1 and oscillator 1.

The power, defined in the same fashion as in \eqref{E:dkhbger},  delivered by the quantum thermal fluctuations of private bath 1 is given by the 1--1 component of the power matrix
\begin{align}
	\mathbf{P}_{\xi}(t)=\operatorname{Re}\langle\mathbf{C}\cdot\bm{\Phi}_{h}(t)\cdot\dot{\mathbf{X}}^{T}(t)\rangle&=\operatorname{Re}\int_{0}^{t}\!ds\;\mathbf{C}\cdot\langle\bm{\Phi}_{h}(t)\cdot\bm{\Phi}_{h}^{T}(s)\rangle\cdot\mathbf{C}\cdot\frac{d}{dt}\mathbf{G}_{R}^{(\chi)T}(t-s)\notag\\
	&=\int_{0}^{t}\!ds\;\mathbf{C}\cdot\mathbf{G}_{H,0}^{(\phi)}(t,s)\cdot\mathbf{C}\cdot\frac{d}{dt}\mathbf{G}_{R}^{(\chi)T}(t-s)\,.
\end{align}
In the late-time limit $t\to\infty$, we obtain
\begin{align}
	\mathbf{P}_{\xi}(\infty)=\int_{-\infty}^{\infty}\!\frac{d\kappa}{2\pi}\;i\kappa\,\mathbf{C}\cdot\widetilde{\mathbf{G}}_{H,0}^{(\phi)}(\kappa)\cdot\mathbf{C}\cdot\widetilde{\mathbf{G}}_{R}^{(\chi)\dagger}(\kappa)\,.
\end{align}
The corresponding power delivered by the nonlocal term in the equation of motion \eqref{E:gbtyer} is
\begin{align}
	\mathbf{P}_{\gamma}(t)&=\operatorname{Re}\int_{0}^{t}\!ds\;\mathbf{C}\cdot\mathbf{G}_{R,0}^{(\phi)}(t-s)\cdot\mathbf{C}\cdot\langle\mathbf{X}(s)\cdot\dot{\mathbf{X}}^{T}(t)\rangle\notag\\
	&=\int_{0}^{t}\!ds\;\mathbf{C}\cdot\mathbf{G}_{R,0}^{(\phi)}(t-s)\cdot\mathbf{C}\cdot\frac{d}{dt}\mathbf{G}_{H}^{(\chi)}(t,s)\,.
\end{align}
In the late time limit, we find
\begin{equation}
	\mathbf{P}_{\gamma}(\infty)=\int_{-\infty}^{\infty}\!\frac{d\kappa}{2\pi}\;i\kappa\,\mathbf{C}\cdot\widetilde{\mathbf{G}}_{R,0}^{(\phi)}(\kappa)\cdot\mathbf{C}\cdot\widetilde{\mathbf{G}}_{H}^{(\chi)}(\kappa)\,.
\end{equation}
Note that the component in the nonlocal term that account for frequency renormalization of the oscillators will not contribute to $\mathbf{P}_{\gamma}$.

The sum of $\mathbf{P}_{\gamma}(\infty)$ and $\mathbf{P}_{\gamma}(\infty)$  will account for the steady flow of thermal energy between the oscillator and its private bath when the dynamics of the oscillator chain reaches an NESS. It is given by 
\begin{align}
	\mathbf{P}_{\xi}(\infty)+\mathbf{P}_{\gamma}(\infty)&=\int_{-\infty}^{\infty}\!\frac{d\kappa}{2\pi}\;i\kappa\Bigl[\mathbf{C}\cdot\widetilde{\mathbf{G}}_{H,0}^{(\phi)}(\kappa)\cdot\mathbf{C}\cdot\widetilde{\mathbf{G}}_{R}^{(\chi)\dagger}(\kappa)+\mathbf{C}\cdot\widetilde{\mathbf{G}}_{R,0}^{(\phi)}(\kappa)\cdot\mathbf{C}\cdot\widetilde{\mathbf{G}}_{H}^{(\chi)}(\kappa)\Bigr]\notag\\
	&=\int_{-\infty}^{\infty}\!\frac{d\kappa}{2\pi}\;\kappa\,\mathbf{C}\cdot\operatorname{Im}\widetilde{\mathbf{G}}_{R,0}^{(\phi)}(\kappa)\cdot\mathbf{C}\cdot\Bigl[\widetilde{\mathbf{F}}(\kappa)\cdot\operatorname{Im}\widetilde{\mathbf{G}}_{R}^{(\chi)}(\kappa)-\widetilde{\mathbf{G}}_{H}^{(\chi)}(\kappa)\Bigr]\,.\label{E:dktbsbs}
\end{align}
From \eqref{E:rnkdfert}, we conclude that since the FDR between  bath 1 and oscillator 1 is not satisfied, the net rate of energy exchange between them will not vanish at late times if the initial temperatures of the two private baths are different. In other words, if there is no initial temperature difference, then there is no thermal current through the oscillator chain no matter how we choose the parameters like $e_{i}$, $\omega_{i}$ and $\sigma$. More importantly, we see the thermal current in the NESS is indeed related to the surplus term in \eqref{E:rnkdfert}. To be more precise, the expressions inside the square brackets in \eqref{E:dktbsbs} give the additional term on the right hand side of \eqref{E:rnkdfert} that prevents one from writing \eqref{E:rnkdfert} into a conventional form of the FDR for an interacting oscillator. We note that it is the difference in the \textit{initial temperatures} of two uncorrelated private baths that matters. The dependence of the system's functions on the initial temperature of the bath field it interacts with is a feature of nonequilibrium dynamics. Since the interaction between the oscillator and its bath is not necessarily weak, each private bath will in general evolve out of its initial thermal state and settle down to a final state. Although this final state barely deviates from its initial thermal state due to the large contrast between the sizes of the phase spaces of the bath field and the oscillator chain, strictly speaking, it will be non-thermal. This deviation will become more significant when the phase space size of the bath gets close to that of the oscillator chain.

\section{Conclusion}\label{S:pturgf}

The fluctuation-dissipation relation may be considered as a categorical relation for any open system which can settle into a stable equilibrium state, in the sense that it is impervious to the details of the system such as the coupling constants. Upon interacting with a thermal bath, a (linear) open system will undergo nonequilibrium evolution from an arbitrary initial state   because its initial state may not be part of the global thermal state. If the system equilibrates, we can identify an FDR for the system in this final equilibrium state, in which the energy exchange between the system and the bath comes into balance. This relaxation process does not depend strongly on the system's initial state, it can be arbitrary, even far from the final equilibrium state. And the system-bath interaction is not restricted to be weak. The FDR in this context is beyond the realm of linear response theory although it has the same familiar form. The difference is, the proportionality factor that equates the noise kernel and the imaginary part of the dissipation kernel of the system depends on the initial temperature of the bath, not on the temperature of the system in the final equilibrium state because the latter cannot be universally introduced for a finite system-bath coupling~\cite{CPR2n4D,MU19}. 

When the system is placed between two thermal baths of different initial temperatures, it also undergoes nonequilibrium evolution. There will be a steady heat flow through the system from one thermal bath to the other. We ask the questions a) whether the system will relax to a steady state, and b) if it does, is there an FDR for this system in NESS. The answers from our present study for this linear model are both affirmative. Comparison to a system interacting with one bath is instructive: In the two bath case, at NESS the relation between the noise and dissipation kernels no longer takes a simple form as in the equilibrium case. An additional term emerges depending on the difference of the initial temperatures of the bath. Its physical meaning can be identified in the expression for the thermal energy flow between the baths in the steady state. Since it is the sole expression in the thermal current that depends on the temperature difference and the coupling strengths of the constituents of the system, it will determine the magnitude of the thermal flow in the nonequilibrium steady state. Knowing the details of what control this heat flow, as our present model study shows, can guide us in the design, and enable one to gain quantitative control in the operation, of thermal devices operating at NESS.\\

\noindent {\bf Acknowledgments} This work is developed when J.-T. H. visited the Maryland Center for Fundamental Physics at the University of Maryland and when B. L. H. visited the Institute of Physics, Academia Sinica and the National Center for Theoretical Sciences in Hsinchu, Taiwan.

\end{document}